\documentclass[twocolumn,prb,showpacs,amsmath,amssymb,english]{revtex4}
\usepackage{graphicx}
\usepackage{dcolumn}
\begin{document}

\title {Variational solution of the $T$-matrix integral equation}

\author{I. A. Nechaev$^{1,2}$ and E. V. Chulkov$^{2,3}$}

\affiliation{$^1$ Tomsk State University, pr. Lenina, 36, 634050 Tomsk, Russia\\
$^2$Donostia International Physics Center (DIPC), P. de Manuel Lardizabal 4,
20018, San Sebasti{\'a}n, Basque Country, Spain\\
$^3$Departamento de F{\'\i}sica de Materiales, Facultad de Ciencias Qu{\'\i}micas, UPV/EHU and Centro Mixto
CSIC-UPV/EHU, Apdo. 1072, 20080 San Sebasti\'an, Basque Country, Spain}

\date{\today}

\begin{abstract}
We present a variational solution of the $T$-matrix integral equation within a local approximation. This
solution provides a simple form for the $T$ matrix similar to Hubbard models but with the local interaction
depending on momentum and frequency. By examining the ladder diagrams for irreducible polarizability, a
connection between this interaction and the local-field factor is established. Based on the obtained
solution, a form for the $T$-matrix contribution to the electron self-energy in addition to the $GW$ term is
proposed. In the case of the electron-hole multiple scattering, this form allows one to avoid double
counting.
\end{abstract}

\pacs{71.10.-w}

\maketitle

\section{Introduction}
As a result of the first cycle of an iterative solution of the Hedin equations,\cite{Hedin} the commonly used
$GW$ approximation (GWA) models the electron self-energy as the product $\Sigma=iG_0W_0$ of a noninteracting
Green function $G_0$ and a dynamically screened Coulomb interaction $W_0$ obtained within the random phase
approximation (RPA). The GWA that describes the long-range screening well has been successfully applied to a
broad spectrum of materials where the interaction is not too strong and screening effects dominate. However,
the GWA encounters difficulties (first of all in its description of the satellite structure) in the case of
systems with localized states where short-range interaction prevails.\cite{FerdiAnis,Solovyev} For such
systems, one has to use a theory beyond the GWA. This theory can be based on both an improvement of the RPA
to get a more realistic screening picture and an inclusion into calculations of the electron self-energy of
the higher-order terms in the screened interaction.

The first attempt to improve the RPA by including the effects of the exchange-correlation (XC) hole is well
known to have been undertaken by Hubbard,\cite{Hubbard} who introduced the so-called local-field factor. The
concept of the latter is that all corrections to the RPA can be formally reduced to it. However, the Hubbard
local-field factor ${\cal G}(\mathbf{q})$ includes the frequency-independent exchange hole correction only.
Diagrammatically such ${\cal G}(\mathbf{q})$ can be exactly derived by summing the ladder diagrams for
irreducible polarizability with a \textit{contact} interaction and noninteracting Green functions (see, e.g.,
Ref.~\onlinecite{Mahan}). In order to explicitly include into consideration the full static XC hole around
the screening electron, Singwi \textit{et al.} \cite{Singwi} have obtained more sophisticated expression for
${\cal G}(\mathbf{q})$ which contains the equilibrium static pair-correlation function.\cite{MahanBook}
Further essential improvements in the derivation of the local-field factor have recently been done by
different authors (see, e.g., Refs. \onlinecite{Richardson,Gusarov,KlausHistory}) who have studied the
frequency dependence of the XC hole.

The concept of the local-field factor has taken on a new physical meaning in time-dependent
density-functional theory (TDDFT).\cite{TDDFT} In the TDDFT within linear response theory, the dynamical
factor ${\cal G}(\mathbf{q},\omega)$ is linked to the XC kernel $f_{xc}(\mathbf{q},\omega)$. The latter plays
the role of the time-dependent (TD) XC interaction in addition to the Coulomb repulsion $v_c$. As a result,
the response function $R$ can be written as \cite{Tokatly}
\begin{equation}\label{Response}
R(q)=P(q)+P(q)v_c(|\mathbf{q}|)R(q),
\end{equation}
where the irreducible polarizability $P$ is defined by the equation
\begin{equation}\label{Polarization}
P(q)=P^0(q)+ P^0(q)f_{XC}(q)P(q).
\end{equation}
Here and in the following we use the four-momentum variable $q$ as a shorthand for $(\mathbf{q},\omega)$. In
Eq.~(\ref{Polarization}) $P^0$ is the RPA irreducible polarizability and $f_{XC}(q)=-v_c(|\mathbf{q}|){\cal
G}(q)$.

In order to derive Eq.~(\ref{Polarization}) from the Hedin equation for the irreducible
polarizability\cite{Hedin}
\begin{equation}\label{irr_pol_Hedin}
P(q)=-\frac{2i}{(2\pi)^4}\int dkG(k)G(k-q)\Lambda(k,q),
\end{equation}
where $G(k)$ is the Green function and $\Lambda(k,q)$ is the vertex function,\cite{remark3poit} the latter
must depend on \textit{one} four-momentum $q$ only (see, e.g., Refs.~\onlinecite{DelSole_94} and
\onlinecite{Hindgren}), i.e.,
\begin{equation}\label{vertex_kernel}
\Lambda(k,q)=\frac{1}{1-f_{XC}(q)P^0(q)},
\end{equation}
which finally leads to $P(q)=P^0(q)\Lambda(q)$.

Diagrammatically such a form for the vertex function has been obtained by Richardson and Ashcroft in
Ref.~\onlinecite{Richardson}, using a \textit{local} approximation\cite{Vignale} within a variational
approach. They have summed an infinite number of self-energy, exchange, and fluctuation terms in the
diagrammatic expansion of $\Lambda$. In contrast to the Hubbard ${\cal G}(\mathbf{q})$, the local-field
factor derived by this summation is a dynamical one.

The representation (\ref{vertex_kernel}) of the vertex function allows one to include vertex corrections into
the calculation of the electron self-energy (see, e.g., Refs.~\onlinecite{Mahan,MahanGWGamma,Hindgren}).
Thus, the concept of the local-field factor suggested by Hubbard considerably simplifies a problem of vertex
corrections calculations in numerical applications and transfers all weight of the problem to calculations of
the local-field factor (or XC kernel) for real systems.

Fundamentally distinct way to go beyond the GWA is based on the use of the $T$
matrix.\cite{FettWal,remark4point} The $T$-matrix approximation (TMA) originally was established to study
strongly correlated fermion systems with short-range interaction and is strictly valid in the limit of an
almost filled or, because of particle-hole symmetry, an almost empty band.\cite{Manghi,Schindlmayr} This
approximation allows one to include processes involving multiple scattering between two electrons or two
holes. This fact makes the TMA capable of describing a satellite structure, for example, in
Ni.\cite{FerdiTM,Manghi,Liebsch,Calandra} However, these calculations were performed using either a
statically screened model interaction \cite{FerdiTM} or the Hubbard parameter $U$ within Hubbard
models.\cite{Schindlmayr,Manghi,Liebsch,Calandra} In the latter, the $T$ matrix in momentum space depends
only on \textit{one} four-momentum [as well as the vertex function (\ref{vertex_kernel}) expressed in terms
of the local-field factor] and schematically can be represented as
\begin{equation}\label{HabbardTmatrix}
T(q)=\frac{U}{1-UK(q)},
\end{equation}
where $K(q)$ is the Fourier transform of the product of two Green functions. In contrast to
Eq.~(\ref{vertex_kernel}), an object of principal concern here is the \textit{local} interaction $U$.

Heuristically combining the simplification of Hubbard models, the $T$-matrix formalism of
Ref.~\onlinecite{FerdiTM}, and a \textit{contact} interaction ${\cal W}={\cal
W}(\mathbf{r},\mathbf{r}';\omega=0)\delta(\mathbf{r}-\mathbf{r}')$ as in Ref.~\onlinecite{Karlsson}, a $GW+T$
matrix approach has recently been developed in Ref.~\onlinecite{Zhukov}. This approach has effectively been
applied to an excited electron lifetime in ferromagnetic Fe and Ni. In fact, comparing with the Hubbard
models, one can find that the model short-range interaction $U$ in the method of Ref.~\onlinecite{Zhukov} is
replaced by the statically screened Coulomb interaction $W_0(\mathbf{q},\omega=0)$. The possibility of such
replacement was recently suggested by several authors.\cite{Springer2,FerdiTM,Ferdi3} Additionally, the
importance of frequency dependence of the Hubbard $U$ has been demonstrated in Ref.~\onlinecite{Ferdi3}.

The motivation of this work is to find a way that allows us to get the same result as the Hubbard model
simplification for the $T$ matrix which is free of model parameters and with the momentum- and
frequency-dependent local screened interaction. In order to accomplish this, we employ a variational
method\cite{Richardson,JhaGuptaWoo} to solve the Bethe-Salpeter equation for the $T$ matrix within a local
approximation. As a result, the $T$ matrix depends only on one four-dimensional wave vector, such as the
vertex function expressed in terms of the local-field factor.

The paper is organized as follows. In Sec.~\ref{sec:t_matrix}, we construct variational functionals and
obtain from the vanishing of their variational derivative a solution of the $T$-matrix integral equation. In
order to connect this solution with the results known from the literature, in Sec.~\ref{sec:polar} we sum the
exchange terms in the diagrammatic expansion of the irreducible polarizability by using the $T$ matrix
obtained. In Sec.~\ref{sec:self_energy} we derive basic formulas for the electron self-energy beyond the GWA.
Finally, the conclusions are given in Sec.~\ref{sec:conclusions}.
\section{\label{sec:t_matrix}$T$ matrix}
\begin{figure}[tbp]
\centering
 \includegraphics[angle=-90,scale=0.7]{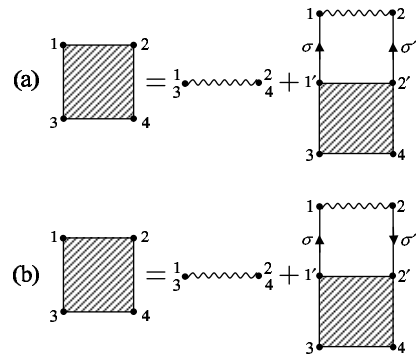}
\caption{Feynman diagrams for $T^{e-e}_{\sigma\sigma'}$ (a) and $T^{e-h}_{\sigma\sigma'}$ (b) in coordinate
space. The $T$ matrix is shown by the shaded square. The wiggly lines signify the dynamically screened
Coulomb interaction $W$. The solid lines with arrows represent the Green function $G$.}\label{tm_all}
\end{figure}

In this section we present mathematical expressions which lead to a simple form for the $T$ matrix depending
on a four-momentum only. We start from the $T$ matrix as an object which will help us in our treatment of the
ladder diagrams both for the irreducible polarizability $P$ and for the electron self-energy $\Sigma$. The
matrix is defined by the following Bethe-Salpeter equation \cite{FettWal,FerdiTM,Zhukov} (Feynman diagrams
are shown in Fig. \ref{tm_all}):
\begin{multline}\label{RSTtype}
    T_{\sigma \sigma '}^{\alpha}(1,2|3,4)=W(1,2)\delta(1-3)\delta(2-4)\\
    +W(1,2)\int\,d1'd2'K_{\sigma \sigma '}^{\alpha}(1,2|1',2')\\
    \times T_{\sigma \sigma '}^{\alpha}(1',2'|3,4),
\end{multline}
where $W$ is the dynamically screened Coulomb interaction and $\sigma$ labels the spin. $\alpha$ can be
specified as $e-e$ in the case of multiple scattering between two electrons or holes and as $e-h$ in the case
of multiple scattering between an electron and a hole. The kernel $K_{\sigma \sigma '}^{\alpha}$ is the
product of the Green functions $G_{\sigma}(1,2)$: 
\begin{eqnarray*}
&&K_{\sigma \sigma'}^{e-e}(1,2|1',2')=iG_{\sigma}(1,1')G_{\sigma '}(2,2'),\\
&&K_{\sigma \sigma'}^{e-h}(1,2|1',2')=iG_{\sigma}(1,1')G_{\sigma '}(2',2).
\end{eqnarray*}
\begin{figure}[tbp]
\centering  \includegraphics[angle=-90,scale=0.7]{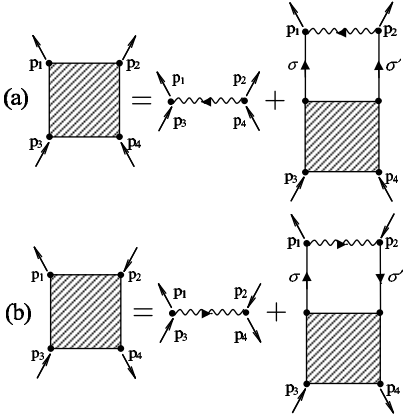}
\caption{Feynman diagrams for $\Gamma^{e-e}_{\sigma\sigma'}$ (a) and $\Gamma^{e-h}_{\sigma\sigma'}$ (b) in
momentum space.}\label{tm_mom_all}
\end{figure}
We have used the shorthand notation $1\equiv (\mathbf{r}_1,t_1)$. As in the majority of practical schemes
(including the commonly used local-density approximation schemes), we suggest for simplicity that the system
considered has properties of a homogeneous system. As a result, the $T$ matrix (\ref{RSTtype}) in momentum
space has the form:\cite{remark_notation}
\begin{eqnarray}\label{MOMTMtype}
T^{\alpha}_{\sigma \sigma'}(p_1,p_2|p_3,p_4)&=&(2\pi)^4\Gamma^{\alpha}_{\sigma\sigma'}(p_1,p_2|p_3,p_4)\nonumber\\
&\times&\delta[p_1\pm p_2-(p_3\pm p_4)].
\end{eqnarray}
In the notations, we use the upper sign for the $e-e$ and the lower sign for the $e-h$ case. The
$\delta$-function in Eq.~(\ref{MOMTMtype}) reflects the conservation of total four-momentum in a homogeneous
system and
\begin{multline}\label{MOMTMtypeGamma}
\Gamma^{\alpha}_{\sigma \sigma'}(p_1,p_2|p_3,p_4)=W(\pm p_1\mp p_3)\\
+\frac{i}{(2\pi)^4}\int\,dkW(k)G_{\sigma}(p_1\mp k)G_{\sigma'}(p_2+k)\\
\times \Gamma^{\alpha}_{\sigma \sigma'}(p_1\mp k,p_2+k|p_3,p_4).
\end{multline}
Feynman diagrams for $\Gamma^{\alpha}_{\sigma \sigma'}$ are shown in Fig. \ref{tm_mom_all}. It is convenient
to introduce the total center-of-mass wave vector and the relative wave vectors \cite{FettWal}
\begin{equation*}
Q=p_1\pm p_2=p_3\pm p_4,\,q=\frac{1}{2}(p_1\mp p_2),\, q'=\frac{1}{2}(p_3\mp p_4).
\end{equation*}
\begin{figure}[tbp]
\centering
 \includegraphics[angle=-90,scale=0.7]{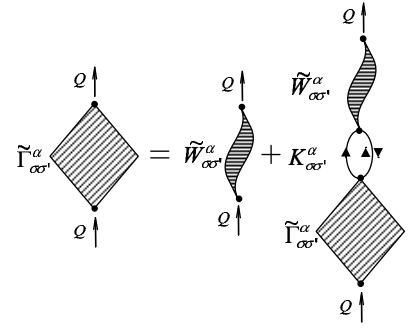}
\caption{A diagrammatic representation of the trial solution $\tilde{\Gamma}^{\alpha}_{\sigma\sigma'}(Q)$,
Eq.~(\ref{TrialSolutionType}), shown for the $e-e$ (up-directed arrow on the right-hand part of the bubble
$K^{\alpha}_{\sigma\sigma'}$) and $e-h$ (down-directed arrow) cases.}\label{trial_solution}
\end{figure}
In terms of these new variables the function $\Gamma^{\alpha}_{\sigma \sigma'}$ from
Eq.~(\ref{MOMTMtypeGamma}) can be cast into the form given by
\begin{multline}\label{Gamma_ne_val_def}
\Gamma^{\alpha}_{\sigma \sigma'}(q,q',Q)\\
\equiv \Gamma^{\alpha}_{\sigma
\sigma'}(\frac{1}{2}Q+q,\pm\frac{1}{2}Q\mp q|\frac{1}{2}Q+q',\pm\frac{1}{2}Q\mp q').
\end{multline}
Defining
\begin{equation}\label{kappa_def}
\kappa^{\alpha}_{\sigma \sigma',Q}(k)=\frac{i}{(2\pi)^4}G_{\sigma}(Q\mp k)G_{\sigma'}(k)
\end{equation}
and
\begin{eqnarray}\label{Phi_kernel}
\Phi_{\sigma \sigma'}^{\alpha}(q,k,Q)&=&\delta(q-k)\nonumber\\
&-&W(\pm q\mp k) \kappa^{\alpha}_{\sigma \sigma',Q}(\pm\frac{1}{2}Q\mp k),
\end{eqnarray}
one derives from the starting equation (\ref{MOMTMtypeGamma}) the relation
\begin{equation}\label{MOMTMtypeGamma3Va}
\int dk\Phi_{\sigma \sigma'}^{\alpha}(q,k,Q)\Gamma_{\sigma \sigma'}^{\alpha}(k,q',Q)=W(\pm q\mp q').
\end{equation}

The integral Eq.~(\ref{MOMTMtypeGamma3Va}) can also be obtained from the vanishing of a functional derivative
\begin{equation}\label{st_cond}
\frac{\delta\mathrm{F}^{\alpha}[G,W,\Gamma]}{\delta\Gamma^{\alpha}_{\sigma\sigma'}(q,q',Q)}=0,
\end{equation}
where $\mathrm{F}$, a functional of three independent variables $G,\,W$, and $\Gamma$, is given by
\begin{multline}\label{typeDirectFunctional}
\mathrm{F}^{\alpha}[G,W,\Gamma]\\
=\sum_{\sigma\sigma'}\int\,dk\,dq'\,dQ\Gamma_{\sigma \sigma'}^{\alpha}(k,q',Q)\kappa^{\alpha}_{\sigma
\sigma',Q}(\pm\frac{1}{2}Q\mp k)\\
\times\bigl\{\int\,dp\,\Phi_{\sigma \sigma'}^{\alpha}(k,p,Q)\Gamma_{\sigma \sigma'}^{\alpha}(p,q',Q)\\
-2W(\pm k\mp q')\bigr\}\kappa^{\alpha}_{\sigma \sigma',Q}(\pm\frac{1}{2}Q\mp q').
\end{multline}
Taking a trial solution in the spirit of the local approximation of Ref.~\onlinecite{Richardson}
\begin{equation}\label{trial_sol}
\Gamma^{\alpha}_{\sigma\sigma'}(q,q',Q)=\tilde{\Gamma}^{\alpha}_{\sigma\sigma'}(Q),
\end{equation}
we find that
\begin{equation}\label{TrialSolutionType}
\tilde{\Gamma}^{\alpha}_{\sigma\sigma'}(Q)=\frac{\widetilde{W}^{\alpha}_{\sigma\sigma'}(Q)}{1-\widetilde{W}^{\alpha}_{\sigma\sigma'}(Q)K^{\alpha}_{\sigma\sigma'}(Q)},
\end{equation}
where
\begin{eqnarray*}
&&K^{\alpha}_{\sigma\sigma'}(Q)=\int dp\,\kappa^{\alpha}_{\sigma \sigma',Q}(p),\\
&&\widetilde{W}^{\alpha}_{\sigma\sigma'}(Q)=
[K^{\alpha}_{\sigma\sigma'}(Q)]^{-1}M^{\alpha}_{\sigma\sigma'}(Q)[K^{\alpha}_{\sigma\sigma'}(Q)]^{-1},\\
&&M^{\alpha}_{\sigma\sigma'}(Q)=\int dq\,dp\,\kappa^{\alpha}_{\sigma
\sigma',Q}(q)W(q-p)\kappa^{\alpha}_{\sigma \sigma',Q}(p).
\end{eqnarray*}

Thus, we have obtained the $T$ matrix as a function of the total center-of-mass wave vector $Q$ only.
Comparing Eq.~(\ref{TrialSolutionType}) with Eq.~(\ref{HabbardTmatrix}), one can see that instead of the
Hubbard parameter $U$ we have a momentum- and frequency-dependent local interaction
$\widetilde{W}^{\alpha}_{\sigma\sigma'}(Q)$. The structure of $\tilde{\Gamma}^{\alpha}_{\sigma\sigma'}$ in
terms of this local interaction is schematically illustrated in Fig.~\ref{trial_solution}.
\section{\label{sec:polar}Irreducible polarizability}
\begin{figure}[tbp]
\centering
 \includegraphics[angle=-90,scale=0.7]{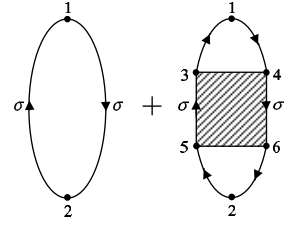}
\caption{Feynman diagrams for the irreducible polarizability $P$ in coordinate space. The RPA bubble (on the
left) and the ladder diagrams (on the right) expressed in terms of the $T$ matrix (shaded square) are
represented here.}\label{pprop}
\end{figure}

We will show here that the $T$ matrix (\ref{TrialSolutionType}) produces the irreducible polarizability in
the form of Eq.~(\ref{Polarization}) with the local-field factor existing in the literature. Actually, the
$T$ matrix allows one to sum the all-order exchange diagrams in the irreducible polarizability diagrammatic
expansion (corresponding Feynman diagrams are shown in Fig. \ref{pprop}):
\begin{eqnarray}\label{PPLTm}
P(1,2)=&&P^0(1,2)\nonumber\\
&&+\sum_{\sigma}\int\,d3d4d5d6G_{\sigma}(1,3)G_{\sigma}(4,1)\nonumber\\
&&\times~ T_{\sigma \sigma}^{e-h}(3,4|5,6)G_{\sigma}(2,6)G_{\sigma}(5,2).
\end{eqnarray}
In momentum space, we have
\begin{eqnarray}\label{MOMPTmGamma3Full}
P(p)=&&-\sum_{\sigma}\int\,dkdq\kappa^{e-h}_{\sigma \sigma,p}(k)\bigl\{\delta(k-q)\nonumber\\
&&+~\Gamma^{e-h}_{\sigma \sigma}(k+\frac{1}{2}p,q+\frac{1}{2}p,p)\kappa^{e-h}_{\sigma \sigma,p}(q)\bigr\}.
\end{eqnarray}
By substituting the $T$ matrix (\ref{TrialSolutionType}) into Eq.~(\ref{MOMPTmGamma3Full}) one
obtains\cite{remark1}
\begin{equation}\label{MOMPTmVarAppr}
P(p)=-\sum_{\sigma}K^{e-h}_{\sigma \sigma}(p)[1-\widetilde{W}^{e-h}_{\sigma\sigma}(p)K^{e-h}_{\sigma
\sigma}(p)]^{-1}.
\end{equation}
As a result, knowing that $P^0(Q)=-\sum_{\sigma}K^{e-h}_{\sigma\sigma}(Q)$, in the local approximation the
irreducible polarizability $P(q)$ for paramagnetic systems has the following familiar form:\cite{remarkTL}
\begin{eqnarray}
P(p)&=&P^0(p)\Lambda(p)\nonumber\\
&=&P^0(p)[1+v_c(\mathbf{q}){\cal G}(p)P^0(p)]^{-1} \label{polFullLocparam}
\end{eqnarray}
with the local-field factor ${\cal G}(p)=\widetilde{W}^{e-h}(p)/2v_c(\mathbf{p})$, where
$\widetilde{W}^{e-h}(p)=\frac{1}{2}\sum_{\sigma}\widetilde{W}^{e-h}_{\sigma \sigma}(p)$.
This factor and the exchange part of the local-field factor of Ref.~\onlinecite{Richardson} are formally the
same.

Next, we notice that, by representing the local interaction as $\widetilde{W}^{e-h}=v_c/\tilde{\varepsilon}$,
the local-field factor can be expressed in terms of the RPA dielectric response function
$\varepsilon^0=1-v_cP^0$ and the first order correction
$\Delta\varepsilon^{(1)}=v_c\sum_{\sigma}M^{e-h}_{\sigma\sigma}$ to $\varepsilon^0$ as\cite{remark2}
\begin{equation}\label{LFFandDRF}
{\cal G}=\frac{1}{2}\tilde{\varepsilon}^{-1}=\frac{\Delta\varepsilon^{(1)}}{[1-\varepsilon^0]^2}.
\end{equation}
A similar expression for the imaginary part of ${\cal G}(q)$ and with the longitudinal Lindhard dielectric
function instead of $\varepsilon^0$ was obtained in Ref.~\onlinecite{Gusarov}, where
$\Delta\varepsilon^{(1)}$ contains the leading corrections to the RPA calculated within the model of the
homogeneous electron gas. At $\omega=0$, the factor (\ref{LFFandDRF}) is akin to the static local-field
factor which has been calculated and parametrized in Ref.~\onlinecite{Kernels2}.

Thus, in the $e-h$ case, we have the transparent connection between the obtained local interaction and the
exchange part of the local-field factor arising from the first order in $W$ exchange irreducible
polarizability diagram. In this sense, the interaction $\widetilde{W}^{e-h}$ agrees conceptually with the XC
kernel considered in Ref.~\onlinecite{TokatlyPankratov}.
\section{\label{sec:self_energy}Self-energy}
\begin{figure}[tbp]
\centering
 \includegraphics[angle=-90,scale=0.7]{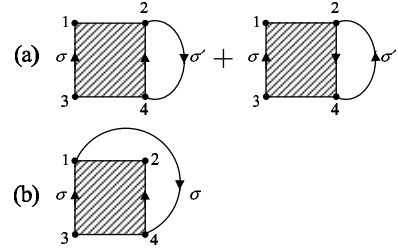}
\caption{Feynman diagrams for the direct (a) and exchange (b) terms of the electron
self-energy.}\label{sigma_all} \end{figure}
In this section we show how the electron self-energy and the $T$ matrix (\ref{TrialSolutionType}) are
related. As is known,\cite{FettWal,FerdiTM,Zhukov} the electron self-energy obtained from the $T$ matrix
consists of a direct term and an exchange one (Feynman diagrams are shown in Fig. \ref{sigma_all}). The
direct term
\begin{multline}\label{SED}
\Sigma_{\sigma}^d(1,3)=
-i\sum_{\sigma'}\int\,d2d4 \bigl\{G_{\sigma'}(4,2)T_{\sigma \sigma '}^{e-e}(1,2|3,4)\\
+G_{\sigma'}(2,4)T_{\sigma \sigma '}^{e-h}(1,2|3,4)\bigr\}
\end{multline}
has $e-e$ and $e-h$ contributions, while the exchange term
\begin{equation}\label{SEEX}
\Sigma_{\sigma}^x(2,3)=i\int\,d1d4 G_{\sigma}(4,1)T_{\sigma \sigma}^{e-e}(1,2|3,4)
\end{equation}
is defined by the spin-diagonal part of the $T^{e-e}$ matrix only.

The Fourier transform of these terms leads to
\begin{multline}\label{MOMSDGamma3}
\Sigma^d_{\sigma}(p)=-\frac{i}{(2\pi)^4}\sum_{\sigma'}\int\,dkG_{\sigma'}(k)\\
\times\bigl\{\Gamma^{e-e}_{\sigma
\sigma'}(\frac{p-k}{2},\frac{p-k}{2},p+k)\\
+\Gamma^{e-h}_{\sigma \sigma'}(\frac{p+k}{2},\frac{p+k}{2},p-k)\bigr\}
\end{multline}
and
\begin{multline}\label{MOMSEXCHGamma3}
\Sigma^x_{\sigma}(p)=\frac{i}{(2\pi)^4}\int\,dkG_{\sigma}(k)\\
\times\Gamma^{e-e}_{\sigma \sigma}(\frac{k-p}{2},\frac{p-k}{2},p+k)
\end{multline}
correspondingly. It is obvious from Eqs. (\ref{MOMSDGamma3}) and (\ref{MOMSEXCHGamma3}) that with the $T$
matrix of Eq.~(\ref{TrialSolutionType}) the exchange term and the spin-diagonal part of the $e-e$
contribution in the direct term are, in fact, identical except for a sign. As a result, as well as in the
Hubbard models, these terms are canceled.

\begin{figure}[tbp]
\centering
 \includegraphics[angle=-90,scale=0.7]{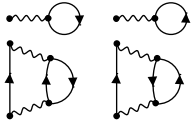}
\caption{Four redundant diagrams originated from the $T^{e-e}$ matrix (left column) and from the $T^{e-h}$
matrix (right column).}\label{diagrams}
\end{figure}

We notice here that, by substituting the $T$ matrix as a solution of Eq.~(\ref{MOMTMtypeGamma3Va}) into Eqs.
(\ref{MOMSDGamma3}) and (\ref{MOMSEXCHGamma3}), one obtains\cite{remark3} for the direct term four lowest
order diagrams (shown in Fig. \ref{diagrams}) which disagree with the solution of the Hedin
equations.\cite{Inkson}
In order to avoid this problem, first of all, following Refs. \onlinecite{FerdiTM} and \onlinecite{Zhukov},
we merely separate the first order exchange term (the GWA electron self-energy term $\Sigma^{GW}_{\sigma}$)
from others. Next, we formally expand the $T$ matrix (\ref{TrialSolutionType}) into series, put into
consideration a new value ${\cal T}^{\alpha}_{\sigma\sigma'}$ containing the second (or third in the $e-h$
case) and higher order in $\widetilde{W}^{\alpha}_{\sigma\sigma'}$ items, and connect this value with the $T$
matrix. This procedure yields
\begin{eqnarray*}
&&{\cal T}^{e-e}_{\sigma\sigma'}(k)=\tilde{\Gamma}^{e-e}_{\sigma \sigma'}(k)K^{e-e}_{\sigma
\sigma'}(k)\widetilde{W}^{e-e}_{\sigma\sigma'}(k),\\
&&{\cal T}^{e-h}_{\sigma\sigma'}(k)=\tilde{\Gamma}^{e-h}_{\sigma \sigma'}(k)[K^{e-h}_{\sigma
\sigma'}(k)\widetilde{W}^{e-h}_{\sigma\sigma'}(k)]^2.
\end{eqnarray*}
On retaining the second order in $\widetilde{W}^{e-e}_{\sigma\sigma'}$ item in ${\cal
T}^{e-e}_{\sigma\sigma'}$, we provide, thereby, the cancellation of the spin-diagonal $e-e$ part of
$\Sigma^{d}_{\sigma}(p)$ and $\Sigma^x_{\sigma}(p)$. Thus, additionally to the $GW$ term, we obtain as a
$T$-matrix contribution to the electron self-energy the following:
\begin{multline}\label{MOMSTLocal}
\Sigma^{T}_{\sigma}(p)=-\frac{i}{(2\pi)^4}\int\,dk\bigl\{G_{-\sigma}(k){\cal T}^{e-e}_{\sigma\,-\sigma}(p+k)\\
+\sum_{\sigma'}G_{\sigma'}(k){\cal T}^{e-h}_{\sigma\sigma'}(p-k)\bigr\}.
\end{multline}
Now we have only one term
\begin{multline}
\Sigma'_{\sigma}(p)=-\frac{i}{(2\pi)^4}\int dk
G_{-\sigma}(k-p)\widetilde{W}^{e-e}_{\sigma\,-\sigma}(k)\\
\times K^{e-e}_{\sigma \,-\sigma}(k)\widetilde{W}^{e-e}_{\sigma\,-\sigma}(k),\nonumber
\end{multline}
which should be excluded from the $T$-matrix contribution (\ref{MOMSTLocal}). As a result, the electron
self-energy can be expressed as $\Sigma_{\sigma}=\Sigma^{GW}_{\sigma}+\Sigma^T_{\sigma}-\Sigma'_{\sigma}$.
The last item is an analog of the so-called double counting term.\cite{FerdiTM,FerdiAnis} In contrast to
Ref.~\onlinecite{Zhukov}, such item is present at the $e-e$ contribution only.
\begin{figure}[tbp]
\centering
 \includegraphics[angle=-90,scale=0.52]{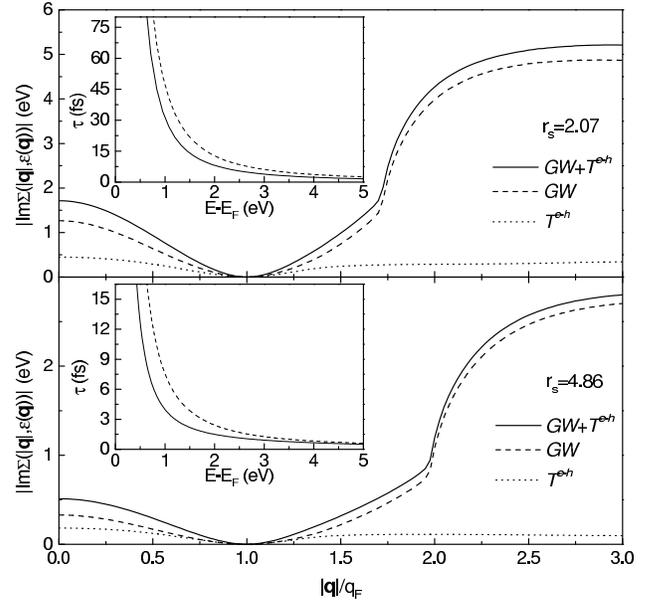}
\caption{The imaginary part of the electron self-energy
$\mathrm{Im}\Sigma[|\mathbf{q}|,\epsilon(\mathbf{q})]$ of the electron gas as a function of momentum
$|\mathbf{q}|$ at $r_s=2.07$ (aluminium) and $r_s=4.86$ (potassium). $\mathrm{Im}\Sigma^{GW}$ and
$\mathrm{Im}\Sigma^{T^{e-h}}$ are shown by dashed and dotted lines, respectively. Solid line represents the
sum of these terms. Insets: the electron lifetime $\tau$ for the corresponding $r_s$ values, versus the
excitation energy $E-E_F$. Dashed (solid) line shows $\tau$ obtained from $\mathrm{Im}\Sigma^{GW}$
($\mathrm{Im}[\Sigma^{GW}+\Sigma^{T^{e-h}}]$). $\epsilon(\mathbf{q})$ is the free electron energy and $q_F$
($E_F$) is the Fermi wave vector (energy).}\label{se_jellium}
\end{figure}

Employing the established connection between $\widetilde{W}^{e-h}(q)$ and ${\cal G}(q)$, one can, in
principle, evaluate the $T^{e-h}$-matrix contribution (\ref{MOMSTLocal}) (denoted as $\Sigma^{T^{e-h}}$) to
the self-energy, additionally to the $GW$ term, by using one of the local-field factors existing in the
literature. But at present it can be seemingly done only for the homogeneous electron gas for which these
factors have been obtained and parametrized.

Here, in order to roughly estimate the magnitude of $\Sigma^{T^{e-h}}$, we exploit the static ${\cal
G}(\mathbf{q})$ of Ref.~\onlinecite{Kernels2}. We have calculated the imaginary part of the electron
self-energy for two values of the electron density corresponding to aluminium ($r_s=2.07$) and potassium
($r_s=4.86$). Our results are shown in Fig.~\ref{se_jellium}. It follows from the figure that in general
$\Sigma^{T^{e-h}}$ is essentially less then $\Sigma^{GW}$ especially in the region where the decay due to
creation of plasmons prevails. However, in the vicinity of the Fermi wave vector the $T^{e-h}$-matrix
contribution amounts on average to $\sim50\%$ ($70\%$) in relation to the $GW$ term for $r_s=2.07$ ($4.86$).
This fact says that the contribution in question can be important in calculations of the decay of excited
electrons whose initial energy is close to the Fermi energy. It is clear from the insets in
Fig.~\ref{se_jellium} that the multiple scattering leads to shortening of the lifetime of such electrons.
Note also that the values of the ratio $\Sigma^{T^{e-h}}/\Sigma^{GW}$ become greater when the electron
density decreases.
\section{\label{sec:conclusions}Conclusions}
In conclusion, we have presented a variational solution of the Bethe-Salpeter equation which determines the
$T$ matrix describing multiple scattering both between two electrons or two holes and between an electron and
a hole. The solution has been obtained within a local approximation. The resulting expression for the $T$
matrix is similar to that in Hubbard models but contains the local interaction depending on momentum and
frequency. Thus the realized variational approach can be viewed as a method to evaluate the local interaction
parameter $U$. In the case of multiple electron-hole scattering, a connection of this interaction with the
local-field factors known from the literature has been established. We have also proposed a form of the
$T$-matrix contribution to the electron self-energy which allows one to sum an infinite number of the
electron-hole ladder diagrams for the electron self-energy without double counting.
\section*{\label{sec:acknowledgments}Acknowledgments}
We thank N.E. Zein for a critical reading of the manuscript. I.A.N. acknowledges V.M. Kuznetsov and V.M.
Silkin for helpful discussions. This work was partially supported by the Research and Educational Center of
Tomsk State University, Departamento de Educaci\'on del Gobierno Vasco, MCyT (Grant No. MAT 2001-0946), and
by the European Community 6th framework Network of Excellence NANOQUANTA (Grant No. NMP4-CT-2004-500198).


\begin{thebibliography}{00}

\bibitem{Hedin} L. Hedin, Phys. Rev. {\bf139}, A796 (1965).

\bibitem{FerdiAnis} F. Aryasetiawan in \emph{Strong Coulomb Correlations in
Electronic Structure Calculations}, edited by V.I. Anisimov (Gordon and Beach, Singapore, 2001).

\bibitem{Solovyev} I.V. Solovyev and M. Imada, Phys. Rev. B {\bf71}, 045103 (2005).

\bibitem{Hubbard} J. Hubbard, Proc. R. Soc. London, Ser. A {\bf243}, 336 (1957).

\bibitem{Mahan} G.D. Mahan, Comments Condens. Matter Phys. {\bf16}, 333 (1994).

\bibitem{Singwi} K.S. Singwi, M.P. Tosi, R.H. Land, and A. Sj\"{o}lander,
Phys. Rev. {\bf176}, 589 (1968).

\bibitem{MahanBook} G.D. Mahan, \emph{Many-Particle Physics} (Plenum Press, New York, 1990).

\bibitem{Richardson} C.F. Richardson and N.W. Ashcroft, Phys. Rev. B {\bf 50}, 8170 (1994).

\bibitem{Gusarov} K. Schturm and A. Gusarov, Phys. Rev. B {\bf62}, 16 474 (2000).

\bibitem{KlausHistory} K. Morawetz, Phys. Rev. B {\bf66}, 075125 (2002).

\bibitem{TDDFT} E. Runge and E.K.U. Gross, Phys. Rev. Lett. {\bf52}, 997 (1984);
E.K.U. Gross and W. Kohn, \emph{ibid}. {\bf55}, 2850 (1985).

\bibitem{Tokatly} I.V. Tokatly, R. Stubner, and O. Pankratov, Phys. Rev. B {\bf65}, 113107 (2002).

\bibitem{remark3poit} In taking the Fourier transform, the vertex function $\Lambda(1,2,3)$ has been
regarding as a function of $1-2$ and $2-3$. Here 1 stands for the four coordinates: space and time,
$1\equiv(\mathbf{r}_1,t_1)$.

\bibitem{DelSole_94} R. Del Sole, L. Reining, and R.W. Godby, Phys. Rev. B {\bf49}, 8024 (1994)

\bibitem{Hindgren} M. Hindgren and C.-O. Almbladh, Phys. Rev. B {\bf56}, 12 832 (1997)

\bibitem{Vignale} G. Vignale and K.S. Singwi, Phys. Rev. B {\bf32}, 2156 (1985).

\bibitem{MahanGWGamma} G.D. Mahan and B.E. Sernelius, Phys. Rev. Lett. {\bf62}, 2718 (1989).

\bibitem{FettWal} A.L. Fetter and J.D. Walecka, \emph{Quantum Theory of Many-particle Systems}
(McGraw-Hill, New York, 1971).

\bibitem{remark4point} In contrast to the vertex function $\Lambda$, the $T$ matrix is a function of the
four space-time coordinates $T(1,2|3,4)$.

\bibitem{Manghi} See, e.g., F. Manghi, V. Bellini, and C. Arcangeli, Phys. Rev. B
{\bf56}, 7149 (1997), and references therein.

\bibitem{Schindlmayr} A. Schindlmayr, T.J. Pollehn, and R.W. Godby,
Phys. Rev. B {\bf58}, 12 684 (1998).

\bibitem{Liebsch} A. Liebsch, Phys. Rev. B {\bf23}, 5203 (1981).

\bibitem{Calandra} C. Calandra and F. Manghi, Phys. Rev. B {\bf45}, 5819 (1992).

\bibitem{FerdiTM} M. Springer, F. Aryasetiawan,  and K. Karlsson, Phys. Rev. Lett.
{\bf80}, 2389 (1998).

\bibitem{Karlsson} K. Karlsson and F. Aryasetiawan, Phys. Rev. B {\bf62}, 3006 (2000).

\bibitem{Zhukov} V.P. Zhukov, E.V. Chulkov, and P.M. Echenique, Phys. Rev. Lett.
{\bf93}, 096401 (2004).

\bibitem{Kernels2} A. Tsolakidis, E.L. Shirley, and R.M. Martin, Phys. Rev. B {\bf69}, 035104 (2004).

\bibitem{Springer2} M. Springer and F. Aryasetiawan, Phys. Rev. B {\bf57}, 4364 (1998).

\bibitem{Ferdi3} F. Aryasetiawan, M. Imada, A. Georges, G. Kotliar, S. Biermann, and A.I. Lichtenstein, Phys.
Rev. B {\bf70}, 195104 (2004).

\bibitem{JhaGuptaWoo} S.S. Jha, K.K. Gupta, and J.W.F. Woo, Phys. Rev. B {\bf4}, 1005 (1971).

\bibitem{remark_notation} Here the $\Gamma$ notation of Ref.~\onlinecite{FettWal} for the Fourier
transform of the $T$ matrix is used.

\bibitem{remark1} It is worth emphasizing that if a trial solution in the form of
$\Gamma^{e-h}_{\sigma\sigma'}(q,q',Q)=\tilde{\Gamma}^{e-h}_{\sigma\sigma'}(q',Q)$ were taken,
Eq.~(\ref{MOMPTmVarAppr}) would be obtained again.

\bibitem{remarkTL} In the case of paramagnetic systems the $T^{e-h}$ matrix of Eq.~(\ref{TrialSolutionType})
can be rewritten as $\Gamma^{e-h}(p)=W^{e-h}(p)\Lambda(p)=2v_c(\mathbf{p}){\cal G}(p)\Lambda(p)$.

\bibitem{remark2} According to Ref.~\onlinecite{Tokatly}, the local-field factor (\ref{LFFandDRF})
and, consequently, $W^{e-h}(p)$ does not contain the $e-h$ singularities presented in the irreducible
polarizability.

\bibitem{TokatlyPankratov} I.V. Tokatly and O. Pankratov, Phys. Rev. Lett. {\bf86}, 2078 (2001).

\bibitem{remark3} This problem does not appear when we use the $T$ matrix starting in its
diagrammatic expansion from the third order in $W$. But in this case the local interaction
$\widetilde{W}^{\alpha}_{\sigma\sigma'}(k)$ has more complicated expression of the third order in $W$.

\bibitem{Inkson} J.C. Inkson, \emph{Many-body Theory of Solids} (Plenum Press, New York, 1984).

\end{thebibliography}
\end{document}